\documentclass[12pt]{article}
\usepackage{graphicx}
\usepackage{amssymb}

\newcommand{\be}{\begin{eqnarray}}
\newcommand{\ee}{\end{eqnarray}}

\begin{document}

\begin{center}

\rightline{LPT Orsay, 04-73} \vspace{0.1cm}
\rightline{RM3-TH/04-22} \vspace{0.1cm}
\rightline{ROME1-1388/2004}

\vspace{1cm}

{\LARGE{\bf SU(3)-breaking effects in kaon and hyperon semileptonic decays from
lattice QCD\footnote{Based on talks given at: {\em DA$\Phi$NE 2004: Physics at
meson factories}, Laboratori Nazionali di Frascati (Italy), June 7-11, 2004; 
{\em VIII International Conference on ``Electron-Nucleus Scattering"}, Marciana
Marina (Italy), June 21-25, 2004; {\em Lattice 2004}, Fermi National Accelerator
Laboratory, Batavia, Illinois (USA), June 21-26, 2004; {\em ICHEP 2004}, Beijing
(China), August 16-22, 2004.}}}

\vspace{1.0cm}

{\large{\sc D.~Be\'cirevi\'c$^1$, D.~Guadagnoli$^2$, G.~Isidori$^3$, 
V. Lubicz$^{4,5}$, 
\\ \vspace{0.1cm}
G.~Martinelli$^2$, F.~Mescia$^{3,4}$, M.~Papinutto$^6$, 
S.~Simula$^5$, 
\\ \vspace{0.2cm}
C.~Tarantino$^{4,5}$ and G.~Villadoro$^2$}}

\vspace{0.6cm}

$^1$LPT, Universit\'e Paris Sud, Centre d'Orsay, F-91405 Orsay-Cedex, France \\
$^2$Dip. di Fisica, Universit\`a di Roma ``La Sapienza'', and INFN, Sezione di 
Roma, \\ P.le A.~Moro 2, I-00185 Rome, Italy \\
$^3$ INFN, Laboratori Nazionali di Frascati, Via E. Fermi 40, I-00044 Frascati, 
Italy \\
$^4$ Dip. di Fisica, Universit\`a di Roma Tre, Via della Vasca Navale 84, 
I-00146 Rome, Italy \\
$^5$ INFN, Sezione di Roma III, Via della Vasca Navale 84, I-00146, Rome, Italy
\\
$^6$ NIC/DESY Zeuthen, Platanenallee 6, D-15738 Zeuthen, Germany

\end{center}

\vspace{1cm}

{\abstract
\noindent
We discuss the result of a recent quenched lattice calculation of the $K\to\pi$
vector form factor at zero-momentum transfer, relevant for the determination of 
$|V_{us}|$ from $K \to \pi \ell \nu$ decays. Using suitable double ratios of 
three-point correlation functions, we show that it is possible to calculate this
quantity at the percent-level precision. The leading quenched effects are 
corrected for by means of quenched chiral perturbation theory. The final result,
$f_+^{K^0\pi^-}(0) = 0.960 \pm 0.005_{\rm stat} \pm 0.007_{\rm syst}$, turns out
to be in good agreement with the old quark model estimate made by Leutwyler and 
Roos. In this paper, we discuss the phenomenological impact of the lattice 
result for the extraction of $|V_{us}|$, by updating the analysis of $K \to \pi 
\ell \nu$ decays with the most recent experimental data. We also present a
preliminary lattice study of hyperon $\Sigma^- \to n \ell \nu$ decays, based on
a similar strategy.}

\newpage

\pagestyle{plain}

\section{Introduction \label{sec:intro}}
Semileptonic decays of kaons and hyperons are of great phenomenological 
interest, as they provide a determination of the CKM matrix element 
$|V_{us}|$~\cite{CKM}. Whereas the most precise estimate of $|V_{us}|$ is
presently obtained from $K \to \pi \ell \nu$ ($K_{\ell3}$) decays, the studies 
of hyperon decays provide an important, independent 
approach~\cite{cabibbo}. Other methods to extract $|V_{us}|$ are based 
on leptonic kaon decays~\cite{milc,marciano-vus} and $\tau$ decays~\cite{tau}.

The analysis of the experimental data on $K \to \pi \ell \nu$ ($K_{\ell3}$) 
\cite{PDG} and hyperon \cite{cabibbo} decays gives access to the quantity 
$|V_{us}| \cdot f_V(0)$, where $f_V(0)$ is the vector form factor at 
zero-momentum transfer for each decay, i.e.~$f_V(0) = f_+(0)$ for $K_{\ell3}$ 
decays and $f_V(0) = f_1(0)$ for hyperon decays. Vector current conservation
guarantees that, in the SU(3)-symmetric limit, $f_+(0)=1$ and $f_1(0)=-1$ for
the $\Sigma^- \to n \ell \nu$ transition. A good theoretical control on these 
decays is obtained via the Ademollo-Gatto (AG) theorem \cite{ag}, which states 
that $f_V(0)$ is renormalized only by terms of at least second order in the 
breaking of the SU(3)-flavor symmetry. The estimate of the difference of 
$f_V(0)$ from its SU(3)-symmetric value represents the main source of 
theoretical error and, in the case of $K_{\ell3}$ decays, it dominates the 
overall uncertainty in the determination of $|V_{us}|$.

The amount of SU(3) breaking due to light quark masses can be investigated in 
the framework of Chiral Perturbation Theory (ChPT), where the form factors can 
be systematically expanded as $f_V(0) = 1 + f_2 + f_3 + \ldots$, with $f_n = 
{\cal{O}}(M^n_K/\Lambda^n_\chi)$ and $\Lambda_\chi \sim$~1~GeV. In the meson 
sector, only the even terms of this expansion appear. Moreover, thanks to the AG
theorem, the first non-trivial term can be computed unambiguously in terms of 
$M_K$, $M_\pi$ and $f_\pi$: $f_2 = -0.023$ in the $K^0 \to \pi^-$ case 
\cite{LR}. At higher orders, the estimate becomes more difficult due to the 
presence of local effective operators whose coupling are essentially unknown. 
The next-to-leading order correction, $f_4$, was evaluated many years ago by 
Leutwyler and Roos \cite{LR}. By employing a general parameterization of the 
SU(3) breaking effects in the meson wave functions, they obtained $f_4 = - 
0.016 \pm 0.008$. This result still represents the reference value adopted by 
the PDG \cite{PDG}. It yields the estimate $f_+(0)=0.961 \pm 0.008$.
 
The complete ChPT calculation of $f_4$ for $K \to \pi$ decays has been recently
performed \cite{BT,post}. The whole result is the sum of a loop amplitude 
(including both genuine two-loop terms and one-loop diagrams with ${\cal O}
(p^4)$ couplings) and a local term that involves a combination of the (unknown) 
${\cal O}(p^6)$ chiral coefficients. The loop amplitude exhibits a large scale 
dependence \cite{CKNRT2}, which is compensated for by the corresponding 
variation of the ${\cal O}(p^6)$ couplings, to yield a scale-independent $f_4$. 
An important observation by Bijnens and Talavera \cite{BT} is that the 
combination of ${\cal O}(p^6)$ constants appearing in $f_4$ could be constrained
in principle by experimental data on the slope and curvature of the scalar form 
factor. The required level of experimental precision, however, is far from what 
can be presently achieved. Thus, for the time being, the only analytic estimates
of the form factor are provided by either the Leutwyler-Roos result or other 
model dependent calculations \cite{CKNRT2,jamin}, and the large scale dependence
of the ${\cal{O}}(p^6)$ loop amplitude \cite{CKNRT2} seems to indicate that the 
$\pm 0.008$ error might be underestimated.

As for hyperon decays, the chiral expansion converges less rapidly (also the odd
$f_i$ are permitted) and the theoretical estimate is quite involved, already at 
the lowest order. The corrections $f_2$ can be unambiguously determined only in 
the limit in which baryons are treated as heavy particles and the octet-decuplet
splitting is treated as a large scale \cite{anderson}. This splitting, however, 
is just of the order of the $K$-meson mass, and it was shown that the 
intermediate decuplet states can partially cancel the (formally leading) octet 
contribution \cite{manohar}. Nothing is known about the higher orders terms. In 
addition, relativistic corrections, proportional to inverse powers of the 
``heavy-baryon" mass $m_B$, have not been investigated so far. Since numerically
$m_B \sim 4\pi f_\pi$, these contributions are expected to be of the same order 
of magnitude of the chiral corrections. Therefore, SU(3)-breaking effects in 
$f_1(0)$ are mainly estimated at present by using phenomenological models 
\cite{models}, which yield corrections of the order of a few percent.

It is clear that a reliable estimate of $|V_{us}|$, from either kaon or
hyperon  semileptonic decays, can be only based on a first principle
determination of the SU(3)-breaking effects. Such a determination can be
provided by lattice QCD, a  non-perturbative approach based only on the
fundamental theory. The first  lattice calculation, with significant accuracy,
of the SU(3)-breaking effects in $K\to\pi$ transitions has been recently
performed \cite{our} (see also  \cite{dafneproc}-\cite{mesciaproc}). A new
strategy has been proposed and successfully applied in the quenched
approximation in order to reach the challenging goal of about 1\% error on
$f_+(0)$.

The aim of this paper is twofold. We first present the quenched lattice result
for $K_{\ell3}$ decays \cite{our} and discuss its impact on the extraction of 
$|V_{us}|$, particularly in the light of the most recent experimental 
determinations. Second, we present the results of an exploratory study of 
hyperon decays based on the same strategy \cite{hyperons}. This study has shown 
that the same level of statistical accuracy reached for $K_{\ell3}$ decays can 
be also achieved for hyperon decays. In this latter case, however, further 
investigations within ChPT are required in order to better control the 
extrapolation to the physical meson masses, necessary in lattice calculations, 
and to reach a level of systematic accuracy comparable to the one achieved in 
the $K_{\ell3}$ case.

\section{Semileptonic $K \to \pi \ell \nu$ decay \label{sec:kaon}}
\subsection{Lattice result and phenomenological implications}

The $K^0 \to \pi^-$ form factors of the weak vector current $V_{\mu} = \bar{s} 
\gamma_{\mu} u$ are defined as
\be
   \langle \pi(p^\prime) | V_{\mu} | K(p)  \rangle = f_+(q^2)(p + 
   p^{\prime})_\mu + f_-(q^2)(p - p^{\prime})_\mu ~, ~~ 
   \label{eq:ff}
\ee 
where $q = p - p^\prime$. As usual, $f_-(q^2)$ can be expressed in terms of the 
scalar form factor
\be
   f_0 (q^2) \equiv f_+(q^2)  +  \frac{q^2}{M_K^2 - M_\pi^2}  f_-(q^2) ~ , 
\ee
so that by construction $f_0(0) = f_+(0)$.

The procedure developed in Ref.~\cite{our} to reach the challenging goal of a 
about 1\% error on $f_+(0)$, is based on the following three main steps:

\begin{itemize}
\item[1)] precise evaluation of the scalar form factor $f_0(q^2)$ at $q^2 = 
q_{\rm max}^2 = (M_K - M_{\pi})^2$;
\item[2)] extrapolation of $f_0(q_{\rm max}^2)$ to $f_0(0) = f_+(0)$;
\item[3)] subtraction of leading chiral logs and extrapolation of $f_+(0)$ to 
the physical meson masses.
\end{itemize}

\noindent 
These steps are described in the next subsections and details of the simulation 
can be found in Ref.~\cite{our}. Here we present our final result for the form 
factor at zero momentum transfer and discuss its phenomenological implications, 
particularly in the light of the most recent experimental results.

By following the procedure outlined above, we obtain the result \cite{our}
\be
   f_+^{K^0\pi^-}(0) = 0.960 \pm 0.005_{\rm stat} \pm 0.007_{\rm syst}\,,
  \label{eq:f0final}
\ee
where the systematic error does not include the estimate of quenched effects 
beyond ${\cal{O}}(p^4)$. Eq.~(\ref{eq:f0final}) compares well with the value 
$f_+^{K^0 \pi^-}(0) = 0.961 \pm 0.008$ quoted by the PDG \cite{PDG} and based on
the Leutwyler-Roos quark model estimate of $f_4$ \cite{LR}, thus putting the
evaluation of the $K\to\pi$ vector form factor on a firmer theoretical basis.
\begin{table}[t]
\centering
\begin{tabular}{r||r||r|r}
& $\delta^K_{SU(2)}$ (\%) 
& \multicolumn{2}{|c}{
$\delta^{K \ell}_{\rm em}(\%) $}    \\
&  &
\multicolumn{2}{|c}{ 3-body $\qquad\qquad$ full}  \\
\hline
$K^{+}_{e3}$    & 2.31 $\pm$ 0.22  &  -0.35  $\pm$ 0.16 &  -0.10 $\pm$  0.16 \\
$K^{0}_{e 3}$   &  0               &  +0.30  $\pm$ 0.10 &  +0.55 $\pm$  0.10 \\
$K^{+}_{\mu 3}$ & 2.31 $\pm$ 0.22  &  -0.05 $\pm$  0.20 &  +0.20 $\pm$  0.20 \\
$K^{0}_{\mu 3}$ &  0               &  +0.55  $\pm$ 0.20 &  +0.80 $\pm$  0.20 \\
\end{tabular}
\caption{\it Summary of the isospin-breaking factors entering
Eq.~(\ref{eq:one}): $\delta^{K \ell}_{\rm em}~ {\rm[3\ body]}$ denotes the 
correction for the inclusive rate excluding the radiative events outside the 
$K_{\ell 3}$ Dalitz Plot \cite{CKNRT2,Cirigliano_old}; $\delta^{K \ell}_{\rm 
em}~{\rm [full]}$ denotes the correction for the fully inclusive rate 
\cite{Andre}. }
\label{tab:iso-brk}
\end{table}

The result in (\ref{eq:f0final}), as well as the  Leutwyler-Roos estimate,
refers to the $K^0 \to \pi^-$ form factor in absence of electromagnetic 
corrections. This quantity is a convenient and well-defined common normalization
for all the four physical photon-inclusive $K_{\ell 3}$ decay widths ($K=\{K^+,
K^0\}$, $\ell=\{e,\mu\}$). By using this definition we can write 
\be
\Gamma(K_{\ell 3}) = { G_F^2 M_K^5 \over 128 \pi^3} |V_{us}|^2  S_{\rm ew}
|f_+^{K^0\pi^-}(0)|^2 C_K^2 I_K^\ell(\lambda_i) 
\left[ 1 + \delta^{K}_{SU(2)} +   \delta^{K \ell}_{\rm em}  \right]^2\,,
\label{eq:one}
\ee
where $G_F=1.1664 \times 10^{-5}$~GeV$^{-2}$, $C_{K}= 1$ ($2^{-1/2}$) for 
neutral (charged) kaon decays, $I_K^\ell(\lambda_i)$ is the kinematical integral
evaluated in the absence of electromagnetic corrections and $S_{\rm ew}=1.0232$ 
is the universal short-distance electromagnetic correction renormalized at the 
scale $\mu=M_\rho$. The residual long-distance component of the electromagnetic 
corrections (which also need to be evaluated at $\mu=M_\rho$ in order to cancel 
the scale dependence of $S_{\rm ew}$) is encoded in $\delta^{K \ell}_{\rm em}$. 
Note that $\delta^{K \ell}_{\rm em}$ depends both on the channel and on possible
experimental cuts applied to the soft-photon radiation. Finally, the strong 
isospin-breaking correction due to $m_u\not=m_d$ is encoded in $\delta^{K}_{SU
(2)}$. The values of these channel-dependent corrections, which can be extracted
by the recent analysis of Refs.~ \cite{CKNRT2,Cirigliano_old,Andre}, are 
reported in Table~\ref{tab:iso-brk}.

\begin{figure}[p]
\begin{center}
\resizebox{0.80\textwidth}{!}{
\includegraphics{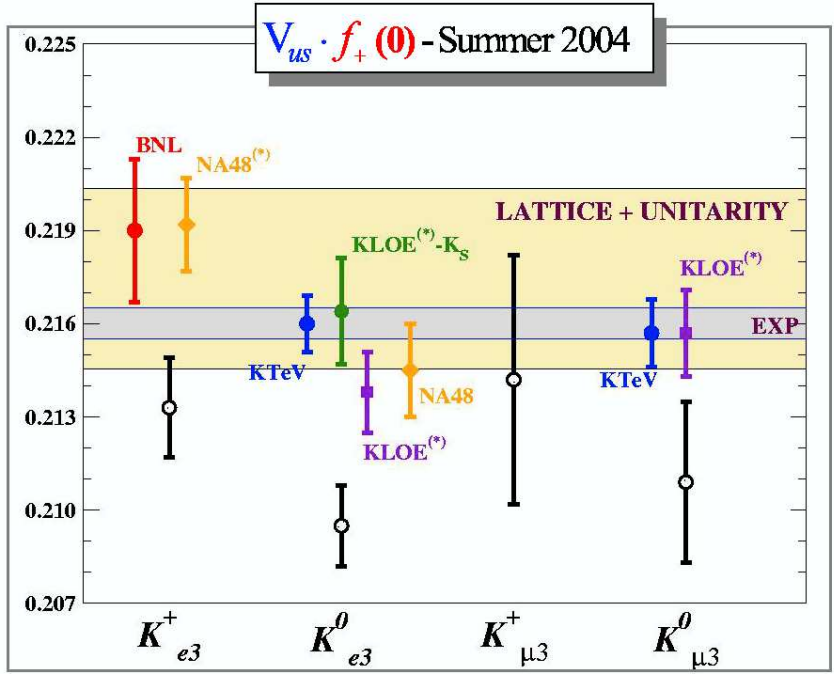}} \vspace{-0.50cm}
\end{center}
\caption{\it Experimental results for $|V_{us}|\cdot f_+(0)$. The gray (darker)
band indicates the average of the new experimental results, $|V_{us}|\cdot 
f_+(0)=0.2160\pm 0.0005$, whereas the yellow (lighter) band represents the 
unitarity prediction combined with our determination of the vector form factor: 
$|V_{us}|^{\mathrm unit.} \cdot f_+(0)=0.2175\pm 0.0029$ (see text). The lower 
(black) points without labels are the old PDG values.}
\label{fig:fvus}
\end{figure}

Inverting Eq.~(\ref{eq:one}) we get 
\be
|V_{us}| \cdot f_{+}^{K^0 \pi^-} (0) = \left[  \frac{ 128 \pi^3 \Gamma^\ell_K}{ 
 G_F^2 M_K^5 S_{\rm ew} C_K^2 I_K^\ell(\lambda_i)  } \right]^{1/2} \, \frac{1}{1
+ \delta^{K}_{SU(2)} +   \delta^{K \ell}_{\rm em} }\,,
\ee 
which could lead to four independent determination of $|V_{us}| \cdot f_{+}^{
K^0 \pi^-} (0)$ from the four $K_{\ell 3}$ modes. Three of these determinations 
have recently been updated by several experiments. A collection of these recent
results is presented in Table~\ref{tab:all} and illustrated in 
Fig.~\ref{fig:fvus}. In this plot we also show the averages of the old results
quoted by the PDG.
\begin{table}[p]
\centering
\begin{tabular}{l||c|c||c|c}
& \multicolumn{2}{c||}{$K^+_{e3}$}
& \multicolumn{2}{c}{$K^L_{\mu3}$} \\
& E865 &  NA48$^{(*)}$ & KTeV & KLOE$^{(*)}$  \\ 
\hline
${\rm BR}$(\%) & $5.13(10)$ & $5.14(6)$  &
$27.01(9)$ & $27.02(25)$ \\ 
\hline
$|V_{us}| \cdot f_{+}^{K^0 \pi^-} (0)$ & $0.2190(23)$ & $0.2192(15)$  & 
$0.2157(11)$ & $0.2157(14)$ \\  
\raisebox{0pt}[15pt][6pt]{}  & \multicolumn{4}{|c}{$K^0_{e3}$}\\
& KTeV[$K_L$]  & \multicolumn{1}{|c|}{NA48[$K_L$]} 
& KLOE[$K_L$]$^{(*)}$   & KLOE[$K_S$]$^{(*)}$ \\ 
\hline
${\rm BR}$(\%) & $40.67(11)$ & \multicolumn{1}{|c|}{$40.10(45)$} & $39.85(35)$  
& $0.0709(11)$ \\ 
\hline
$|V_{us}|  \cdot f_{+}^{K^0 \pi^-} (0)$ & $0.2160(9)$ & \multicolumn{1}{|c|}
{$0.2145(15)$} & $0.2138(13)$  & $0.2164(17)$ \\  
\end{tabular}
\caption{\em Summary of recent results on $K_{\ell 3}$ decays obtained by 
BNL-E865 \cite{E865}, KTeV \cite{KTeV},  NA48 \cite{NA48} and KLOE \cite{KLOE}
and corresponding values of $|V_{us}|\cdot f_+(0)$. The results marked by $(*)$ 
are still preliminary. The values of $|V_{us}| f_{+}^{K^0 \pi^-} (0)$ have been 
computed using the PDG values for the lifetimes \cite{PDG} and using, in all 
channels, the slopes fitted by KTeV with a polar fit~\cite{KTeV_slope}. 
\label{tab:all}}
\end{table}
The new data consistently point toward a value of $|V_{us}| \cdot f_{+}^{K^0 
\pi^-} (0)$  substantially larger with respect to the one derived by the old PDG
results. The value of $|V_{us}|$ extracted from these new data, combined with 
our estimate of $f_{+}^{K^0 \pi^-} (0)$, is
\be
|V_{us}| = 0.2250 \pm 0.0020_{f_+(0)} \pm 0.0005_{\rm exp} =  0.2250 \pm 0.0021
\,.
\ee
Interestingly, this result is perfectly consistent with the value derived by CKM
unitarity, thus solving a long-standing problem (see e.g.~Ref.~\cite{GI}). Using
the accurate determination of $|V_{ud}|$ from nuclear $0^+\to0^+$ and nucleon 
beta decays, $|V_{ud}| = 0.9740 \pm 0.0005$ \cite{marciano}, one finds indeed 
$|V_{us}|^{\rm unit.}\simeq \sqrt{1-|V_{ud}|^2} = 0.2265 \pm 0.0022$.

Given our close agreement with the Leutwyler-Roos result, the solution of the 
CKM unitarity problem perfectly holds also if their estimate of $f_{+}^{K^0 \pi
^-} (0)$ is adopted. On the other hand, the consistency is substantially worse 
with the value $f^{K^0 \pi^-}(0)=0.976\pm 0.010$ of Ref.~\cite{BT}. It is also
worth to mention that the estimate $f^{K^0 \pi^-}(0)=0.981\pm 0.010$ of 
Ref.~\cite{CKNRT2} is not independent from the one of Ref.~\cite{BT}, and 
{\em should not} be used in conjunction with Eq.~(\ref{eq:one}) and/or compared 
with our result. The difference between Ref.~\cite{CKNRT2} and most other 
analyses is due to a {\em partial inclusion} of the electromagnetic corrections 
in the form factor (concerning SU(3) breaking effects, they simply use the 
results of Ref.~\cite{BT}). In our treatment, as well as in the analysis 
performed by the KTeV Collaboration \cite{KTeV}, all the electromagnetic 
corrections are explicitly factorized out. This approach has the advantage of 
separating the hard theoretical problem of estimating SU(3) breaking effects 
(identical for the four decay modes), from the less severe (but 
channel-dependent) issue of estimating  electromagnetic corrections.

\subsection{Evaluation of $\mathbf{f_0(q^2)}$ at $\mathbf{q^2 = q_{\rm max}^2 
= (M_K - M_{\pi})^2}$}

In the next subsections we discuss the strategy of the lattice calculation of 
the vector form factor. A more exhaustive discussion and all details of the
numerical simulation can be found in Ref. \cite{our}.

By following a procedure originally proposed in Ref.~\cite{FNAL} to study 
heavy-light form factors, the scalar form factor $f_0(q^2)$ can be calculated 
very efficiently at $q^2 = q_{\rm max}^2 = (M_K - M_\pi)^2$ (i.e.~$\vec p = 
\vec{p}^{\,\prime} = \vec q = 0$) from the double ratio of three-point 
correlation functions with both mesons at rest:
\be
   R_0(t_x, t_y) \equiv \frac{C_0^{K \pi}(t_x,t_y,\vec 0,\vec 0) \, 
   C_0^{\pi K}(t_x,t_y,\vec 0,\vec 0)} {C_0^{K K}(t_x,t_y,\vec 0,\vec 0) \, 
   C_0^{\pi \pi}(t_x,t_y,\vec 0,\vec 0)} \,,
   \label{eq:fnal1}
\ee
where the three-point correlation function for the $K \to \pi$ transition is 
defined as
\be
   C_\mu^{K \pi} (t_x,t_y,\vec p,\vec{p}^{\,\prime}) = \sum_{\vec x, 
   \vec y} \langle O_\pi(t_y,\vec y) ~ \widehat{V}_\mu(t_x,\vec x) ~ 
   O_K^\dagger(0) \rangle \cdot e^{-i \vec p \cdot \vec x + 
   i \vec{p}^{\,\prime} \cdot (\vec x - \vec y)}
   \label{eq:c3pt}
\ee
with $\widehat{V}_{\mu}$ being the renormalized lattice weak vector current and 
$O_\pi^{\dagger} = \bar d \gamma_5 u$, $O_K^{\dagger} = \bar d \gamma_5 s$ the 
meson interpolating fields.

When the vector current and the two interpolating fields are separated far 
enough from each other, the contribution of the ground states dominates the
correlation functions, yielding
 \be 
    R_0(t_x, t_y)_{\overrightarrow{\stackrel{\mbox{\tiny $t_x \to 
    \infty$}}{\mbox{\tiny $(t_y - t_x) \to \infty$}}}} \frac{\langle \pi 
    | \bar{s} \gamma_0 u | K \rangle \, \langle K | \bar{u} \gamma_0 s | 
    \pi \rangle}{\langle K | \bar{s} \gamma_0 s | K \rangle \, \langle \pi | 
    \bar{u} \gamma_0 u | \pi \rangle} = [f_0(q_{\rm max}^2)]^2 \,\frac{(M_K +
    M_\pi)^2}{4 M_K M_\pi}\,.
    \label{eq:fnal2}
\ee
Thus from this ratio the scalar form factor at maximum $q^2$ can be directly
determined.

There are several crucial advantages in the use of the double ratio 
(\ref{eq:fnal1}). First, there is a large reduction of statistical 
uncertainties, because fluctuations in the  numerator and the denominator are 
highly correlated. Second, the matrix elements of the meson sources cancel 
between numerator and denominator. Third, the double ratio is independent from 
both the renormalization constant and the ${\cal{O}}(a)$-improvement coefficient
of the vector current, where $a$ is the lattice spacing. Therefore $R_0$ suffers
from discretization effects only at order $a^2$. Finally, the double ratio is 
equal to unity in the SU(3)-symmetric limit at all orders in $a$. Thus the 
deviation of $R_0$ from unity depends on the physical SU(3) breaking effects on 
$f_0(q_{\rm max}^2)$ as well as on discretization errors, but the latter are at 
least of order $a^2 (m_s - m_\ell)^2$. In other words, our result for $f_0(q_
{\rm max}^2)$ is not affected by the whole discretization error on the 
three-point correlation function, but only by its smaller SU(3)-breaking part. 
Similar considerations apply also to the quenching error, because the double 
ratio $R_0(t_x, t_y)$ is correctly normalized to unity in the SU(3)-symmetric 
limit also in the quenched approximation.

\begin{figure}[htb]
\begin{center}
\resizebox{0.60\textwidth}{!}{
\includegraphics{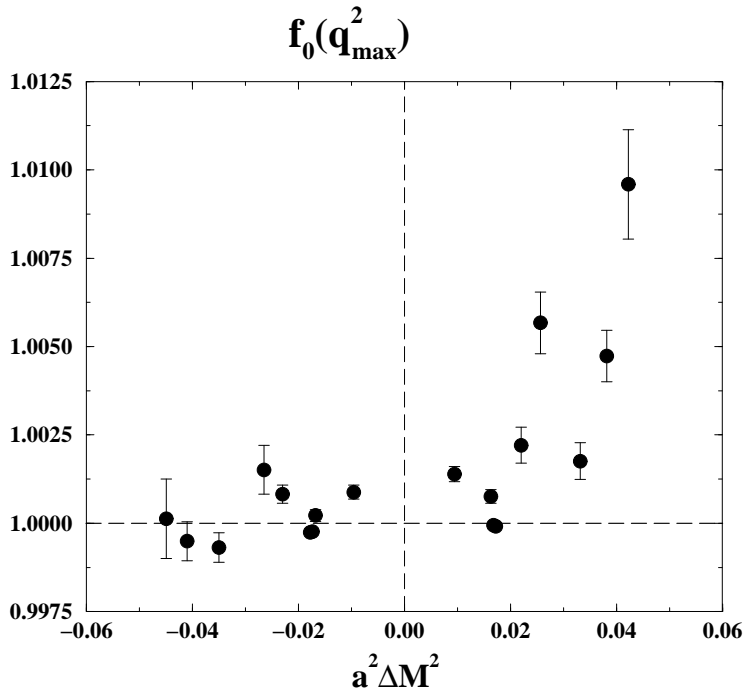}} \vspace{-0.5cm}
\end{center}
\vspace{-0.5cm}
\caption{\it Values of $f_0(q_{\protect\rm max}^2)$, obtained in 
Ref.~\protect\cite{our} using Eq.~(\protect\ref{eq:fnal2}), versus the 
SU(3)-breaking parameter $a^2 \Delta M^2 \equiv a^2 (M_K^2 - M_{\pi}^2)$.}
\label{fig:f0qmax}
\end{figure}

From the plateau of the double ratio~(\ref{eq:fnal2}) the values of $f_0(q_{\rm 
max}^2)$ are determined with a statistical uncertainty smaller than 0.1\%, as it
is illustrated in Fig.~\ref{fig:f0qmax}.

\subsection{Extrapolation of $\mathbf{f_0(q_{\rm max}^2)}$ to $\mathbf{f_0(0) = 
f_+(0)}$}
The extrapolation of $f_0(q^2)$ from $q_{\rm max}^2$ to $q^2 = 0$ requires the 
knowledge of the slope of $f_0$, and thus the study of the $q^2$-dependence of 
the scalar form factor. An important remark is that, in order to get $f_0(0)$ at
the percent level, the precision required for the slope can be much lower 
($\simeq 30$\%), because the values of $q_{\rm max}^2$ in the numerical
simulation can be chosen very close to $q^2 = 0$.

For each set of quark masses two- and three-point correlation functions have
been calculated for mesons with various momenta in order to extract the $q^2$ 
dependence of both $f_0(q^2)$ and $f_+(q^2)$. The latter turns out to be well 
determined on the lattice with a statistical error of about $5 \div 20$\% and 
its $q^2$-dependence is very well described by a pole-dominance fit, $f_+(q^2) 
= f_+(0) / [1 - \lambda_+ ~ q^2]$. On the contrary, for the scalar form factor 
the uncertainties are about $5$ times larger. As explained in Ref.~\cite{our} 
the precision in the extraction of $f_0(q^2)$ can be significantly improved by 
constructing new suitable double ratios 
\be
   R_i(t_x, t_y) = \frac{C_i^{K \pi}(t_x, t_y, \vec p, \vec{p}^{\,\prime})} 
   {C_0^{K \pi}(t_x, t_y, \vec p, \vec{p}^{\,\prime})} ~ \frac{C_0^{K K}(t_x, 
   t_y, \vec p, \vec{p}^{\,\prime})}{C_i^{K K}(t_x, t_y, \vec p, 
   \vec{p}^{\,\prime})} ~ ,
   \label{eq:rappd}
\ee
(with $i=1,2,3$) from which a determination of the ratio of the form factors 
$f_0(q^2)/f_+(q^2)$ is obtained. The advantages of the double ratios 
(\ref{eq:rappd}) are similar to those already pointed out for the double ratio 
(\ref{eq:fnal1}), namely: i) a large reduction of statistical fluctuations; ii) 
the independence of the improved renormalization constant of the vector current,
and iii) $R_i \to 1$ in the SU(3)-symmetric limit. We stress that the 
introduction of the matrix elements of degenerate mesons in Eq.~(\ref{eq:rappd})
is crucial to largely reduce statistical fluctuations, because it compensates 
the different fluctuations of the matrix elements of the spatial and time 
components of the weak vector current.

\begin{figure}[t]
\begin{center}
\resizebox{0.60\textwidth}{!}{
\includegraphics{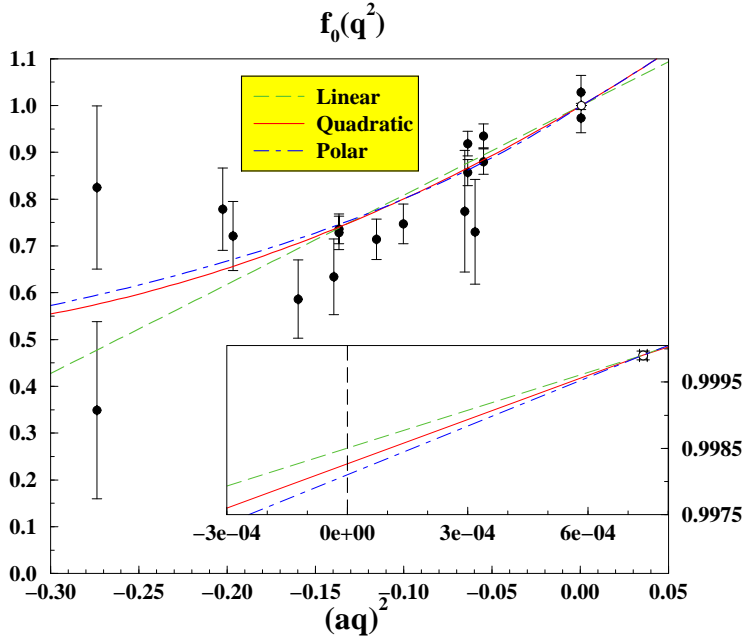}} \vspace{-0.50cm}
\end{center}
\vspace{-0.5cm}
\caption{\it The form factor $f_0(q^2)$ obtained from the double ratios 
(\ref{eq:rappd}) for $q^2 < q_{\rm max}^2$ (full dots) and from the double 
ratio (\ref{eq:fnal1}) at $q^2 = q_{\rm max}^2$ (open dot), for the quark mass 
combination $k_s = 0.13390$ and $k_\ell = 0.13440$. The dot-dashed, dashed and 
solid lines correspond to the polar, linear and quadratic fits given in 
Eqs.~(\ref{eq:polar}-\ref{eq:quadratic}), respectively. The inset is an 
enlargement of the region around $q^2 = 0$.}
\label{fig:f0fit}
\end{figure}

Thanks to the ratios (\ref{eq:rappd}) the statistical uncertainties on 
$f_0(q^2)$ become about $5 \div 20$\%. The quality of the results is shown in 
Fig.~\ref{fig:f0fit} for one of the combinations of quark masses used in 
Ref.~\cite{our}. The points are paired because both $K \to \pi$ and $\pi \to K$ 
transitions are considered.

In order to extrapolate the scalar form factor to $q^2 = 0$ we have considered 
three different possibilities, namely a polar, a linear and a quadratic fit:
\be
   f_0(q^2) & = & f_0^{(pol.)}(0) / (1 - \lambda_0^{(pol.)} \, q^2) \,,  
   \label{eq:polar} \\
   f_0(q^2) & = & f_0^{(lin.)}(0) \cdot (1 + \lambda_0^{(lin.)} \, q^2) \,, 
   \label{eq:linear} \\
   f_0(q^2) & = & f_0^{(quad.)}(0) \cdot (1 + \lambda_0^{(quad.)} \, q^2 + 
   c_0 \, q^4) \,.
   \label{eq:quadratic}
\ee
These fits are shown in Fig.~\ref{fig:f0fit} and provide values of both $f_0(0)$
and the slope $\lambda_0$, which are consistent with each other within the 
statistical uncertainties.

Lattice artifacts on $f_0(0)$, due to the finiteness of the lattice spacing, 
start at ${\cal{O}}(a^2)$ and are proportional to $(m_s - m_\ell)^2$, like the 
physical SU(3)-breaking effects. Indeed, the determination of 
$f_0(q_{\rm max}^2)$ is affected only by discretization errors of ${\cal{O}}
[a^2 (m_s - m_\ell)^2]$, because the double ratio (\ref{eq:fnal1}) is 
${\cal{O}}(a)$-improved and symmetric with respect to the exchange $m_s 
\leftrightarrow m_\ell$ in the weak vertex. In addition, since $q_{\rm max}^2$ 
is proportional to $(m_s - m_\ell)^2$, ${\cal{O}}(a^2)$ effects in the 
extrapolation from $q_{\rm max}^2$ to $q^2 = 0$ also vanish quadratically in 
$(m_s - m_\ell)$. Being in our calculation $a^{-1} \simeq 2.7$ GeV, we expect 
discretization errors to be sensibly smaller than the physical 
SU(3)-breaking effects. 

\begin{figure}[t]
\begin{center}
\resizebox{0.60\textwidth}{!}{
\includegraphics{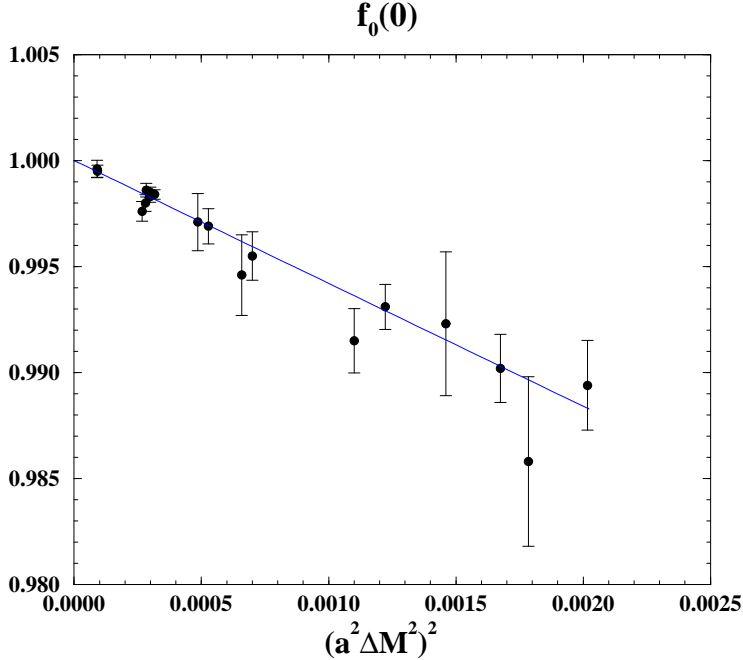}} \vspace{-0.50cm}
\end{center}
\vspace{-0.5cm}
\caption{\it Values of $f_0(0) = f_+(0)$, obtained from the quadratic fit 
(\ref{eq:quadratic}), versus $(a^2 \Delta M^2)^2$. The solid line is the result 
of the linear fit $f_0(0) = 1 - A ~ (a^2 \Delta M^2)^2$ where $A$ is a 
mass-independent parameter.}
\label{fig:plotf0}
\end{figure}

The results obtained for $f_0(0)$ agree well with a quadratic dependence on 
$a^2 \Delta M^2$, where $\Delta M^2=M_K^2-M_\pi^2$, as expected from both 
physical and lattice artifact contributions. This feature is clearly visible in 
Fig.~\ref{fig:plotf0}, where the values of $f_0(0)$ obtained from the quadratic 
fit (\ref{eq:quadratic}), are plotted.

\subsection{Results for the slopes}
Our results for the slope $\lambda_0$, extrapolated to the physical meson masses
(using a linear dependence in the quark masses) and given in units of $M_{\pi^+}
^2$, yield: $\lambda_0^{(pol.)} = 0.0122(22)$, $\lambda_0^{(lin.)} = 0.0089(11)$
and $\lambda_0^{(quad.)} = 0.0115(26)$. Our ``polar" value $\lambda_0^{(pol.)}$ 
is consistent with the recent determination from KTeV $\lambda_0^{(pol.)} = 
0.01414 \pm 0.00095$ \cite{KTeV_slope} and represents a true theoretical 
prediction, having been obtained before the KTeV result were published. 

As far as the vector form factor is concerned, the values obtained for its
slope $\lambda_+$ (employing the polar parametrization) agree well with the
inverse of the squared $K^*$-meson mass, for each combination of the simulated
quark  masses. A simple linear extrapolation in terms of the quark masses to
the  physical values yields $\lambda_+ =  0.026 \pm 0.002$ in units of
$M_{\pi^+}^2$, which is consistent with the PDG value $\lambda_+ =  0.028 \pm
0.002$ \cite{PDG} (obtained from linear fits to the vector form factor). The
agreement is even better with the recent KTeV result obtained using a polar
fit: $\lambda_+ = 0.02502 \pm 0.00037$  \cite{KTeV_slope}. Accurate
determinations of the slopes $\lambda_+$ and $\lambda_0$  have recently been
reported also by ISTRA+ \cite{ISTRA+} and NA48 ($\lambda_+$ only)
\cite{NA48_slope}. The results obtained by means of linear parameterizations
are in good agreement with those obtained by KTeV. The situation of the
quadratic fit to the vector form factor is more controversial. For the polar
form, which is the one suggested also by the lattice calculation (that explores 
a large range in $q^2$), the NA48 result is in agreement with the one obtained 
by KTeV, whereas this parameterization has not been investigated by ISTRA+. The
choice of the polar form seems to be the best compromise among the various 
results, and for this reason it has been used in extracting the experimental 
values of $|V_{us}| \cdot f_{+}(0)$ presented in Table \ref{tab:all}.

\subsection{Extrapolation of $\mathbf{f_+(0)}$ to the physical masses}
\begin{figure}[t]
\begin{center}
\resizebox{0.60\textwidth}{!}{
\includegraphics{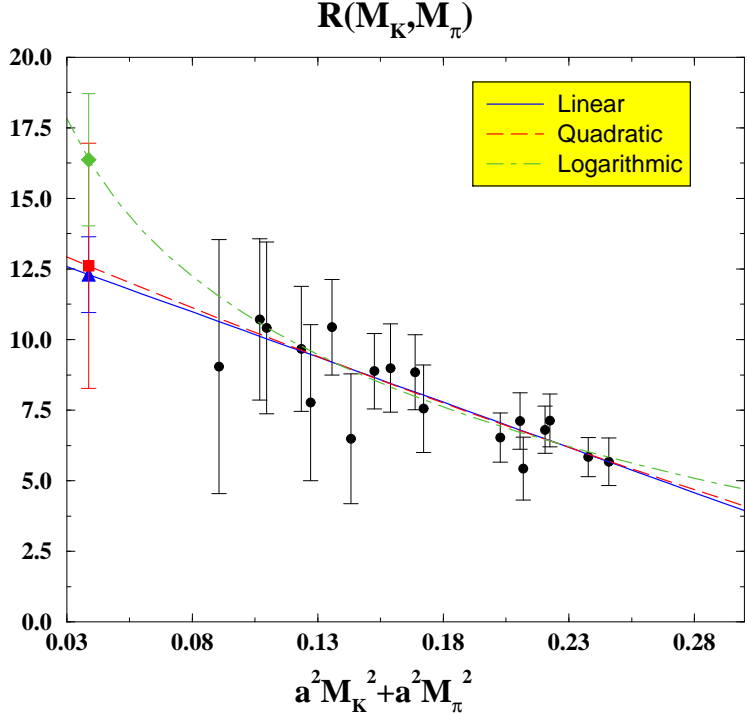}} \vspace{-0.50cm}
\end{center}
\vspace{-0.5cm}
\caption{\it Comparison among linear (\ref{eq:agfit}), quadratic 
(\ref{eq:qmfit}) and logarithmic (\ref{eq:qlfit}) fits of the ratio 
$R(M_K, M_\pi)$ as a function of $[a^2 M_K^2 + a^2 M_{\pi}^2]$. Triangle, 
square and diamond are the values of $R(M_K, M_\pi)$ extrapolated to the 
physical meson masses. For illustrative purposes we have chosen the case in 
which a quadratic fit in $q^2$is used to extrapolate the scalar form factor to 
$q^2 = 0$.}
\label{fig:fitcfr}
\end{figure}

In order to determine the physical value of $f_+(0)$ the lattice results of 
Fig.~\ref{fig:plotf0} are extrapolated to the physical kaon and pion masses. The
problem of the chiral extrapolation is substantially simplified if the AG 
theorem (which also holds in the quenched approximation) is taken into account 
and if the leading (quenched) chiral logs are subtracted. Thus in \cite{our} the
following quantity is introduced
\be
   R(M_K, M_\pi) =  \frac{ 1 + f_2^q(M_K, M_\pi) -f_+(0; M_K, M_\pi) }{(a^2 
   \Delta M^2)^2} \,,
   \label{eq:linfit}
\ee
where $f_2^q$ represents the leading non-local contribution, determined by 
pseudoscalar meson loops within quenched ChPT, calculated in Ref.~\cite{our}. 
The subtraction of $f_2^q$ is a well defined procedure being finite and 
scale-independent. It should be emphasized that this subtraction in 
Eq.~(\ref{eq:linfit}) does not imply necessarily a good convergence of 
(quenched) ChPT at ${\cal O}(p^4)$ for the meson masses used in the lattice 
simulation. The aim of this procedure is to access directly on the lattice the 
quantity $1 + f_2^q-f_+(0)$, defined in such a way that its chiral expansion 
starts at ${\cal O}(p^6)$, independently of the values of the meson masses. 
After the subtraction of $f_2^q$ we expect that, in the range of masses 
considered in the lattice simulation, the ratio $R(M_K, M_\pi)$ receives large 
contributions from ${\cal O}(p^6)$ local operators in the effective theory. At 
the same time, the quadratic dependence on $a^2 \Delta M^2$, driven by the AG 
theorem, is already factorized out. Hence $R(M_K, M_\pi)$ should be a quantity 
well suited for a smooth polynomial extrapolation in the meson masses. It turns 
out, indeed, that the dependence of $R(M_K, M_\pi)$ on the meson masses is well 
described by a simple linear fit:
\be
   R^{(lin.)}(M_K, M_\pi) = c_{11} + c_{12} [(a M_K)^2 + (a M_\pi)^2] ~ ,
   \label{eq:agfit}
\ee
whereas the dependence on $\Delta M^2$ is found to be negligible. In order to 
check the stability of the results, quadratic and logarithmic fits have been 
also considered:
\be
   R^{(quad.)}(M_K, M_\pi) & = & c_{21} + c_{22} [(a M_K)^2 + (a M_\pi)^2] + 
   c_{23} [(a M_K)^2 + (a M_\pi)^2]^2 , ~~ 
   \label{eq:qmfit} \\
   R^{(log.)}(M_K, M_\pi) & = & c_{31} + c_{32} \log[ (a M_K)^2 + 
   (a M_\pi)^2] . 
   \label{eq:qlfit}
\ee
In Fig.~\ref{fig:fitcfr} it is shown that linear, quadratic and logarithmic 
functional forms provide equally good fits to the lattice data with consistent 
results also at the physical point.

Combining our estimate of $R(M_K, M_\pi)$ at the physical meson masses with the 
unquenched value of $f_2$ ($f_2 = -0.023$ \cite{LR}), we finally obtain the
result
\cite{our}
 \be
    f_+^{K^0\pi^-}(0) = 0.960 \pm 0.005_{\rm stat} \pm 0.007_{\rm syst}\,,
   \label{eq:f0final2}
 \ee
which has been anticipated in Eq.~(\ref{eq:f0final}). Note that the systematic 
error does not include an estimate of quenched effects beyond ${\cal{O}}(p^4)$.
Removing this error represents one of the major goal of future lattice studies
of $K_{\ell 3}$ decays. The error quoted in Eq.~(\ref{eq:f0final2}) is mainly 
due to the uncertainties resulting from the functional dependence of the scalar 
form factor on both $q^2$ and the meson masses. It can be further reduced by 
using larger lattice volumes (leading to smaller lattice momenta) as well as 
smaller meson masses. We stress again that in our estimate of $f_+(0)$ 
discretization effects start at ${\cal{O}}(a^2)$ and are also proportional to 
$(m_s - m_\ell)^2$, as the physical SU(3)-breaking effects. Thus our result is 
not affected by the whole discretization error on the three-point correlation 
function, but only by its smaller SU(3)-breaking part. Discretization errors on 
$\left[f_+(0)-1-f_2 \right]$ are estimated to be few percent of the physical 
term, i.e.~well within the systematic uncertainty quoted in 
Eq.~(\ref{eq:f0final2}). For a more refined estimate of these effects, 
calculations at different values of the lattice spacing are required.

\section{Semileptonic $\mathbf{\Sigma^- \to n \ell \nu}$ decay\label{sec:sigma}}
Semileptonic hyperon decays represent the ``baryonic way'' to a precise
determination of $|V_{us}|$. In this Section we present the results of a 
preliminary lattice study of the vector form factor $f_1(0)$ for the $\Sigma^- 
\to n \ell \nu$ decay, based on a strategy similar to the one developed for kaon
decays. More details on this calculation can be found in Ref.~\cite{hyperons}.

The relevant matrix element of the weak vector current can be decomposed in 
terms of the following structures 
\be
    <n(p^{\prime})| \overline{u} \gamma^\mu s |\Sigma^-(p)> = 
    \overline{u}_n(p^{\prime}) 
    \left\{ \gamma^\mu f_1(q^2) - \frac{i \sigma^{\mu \nu} q_\nu}{M_n + 
    M_\Sigma} f_2(q^2) + \frac{q^\mu}{M_n + M_\Sigma} f_3(q^2) \right\} 
    u_\Sigma(p) ~~ 
\ee
with $q = p - p^{\prime}$. As in the case of $K_{\ell3}$ decays, one introduces 
the scalar form factor $f_0(q^2)$:
\be
   f_0(q^2) = f_1(q^2) + \frac{q^2}{M_\Sigma^2 - M_n^2} ~ f_3(q^2) ~,
\ee
so that $f_0(0) = f_1(0)$. Note that the SU(3)-symmetric value of $f_1(0)$ is 
given by a Clebsch-Gordan coefficient, equal to ($-1$) for the $\Sigma^- \to n$ 
transition.

\begin{figure}[t]
\begin{center}
\resizebox{0.70\textwidth}{!}{
\includegraphics{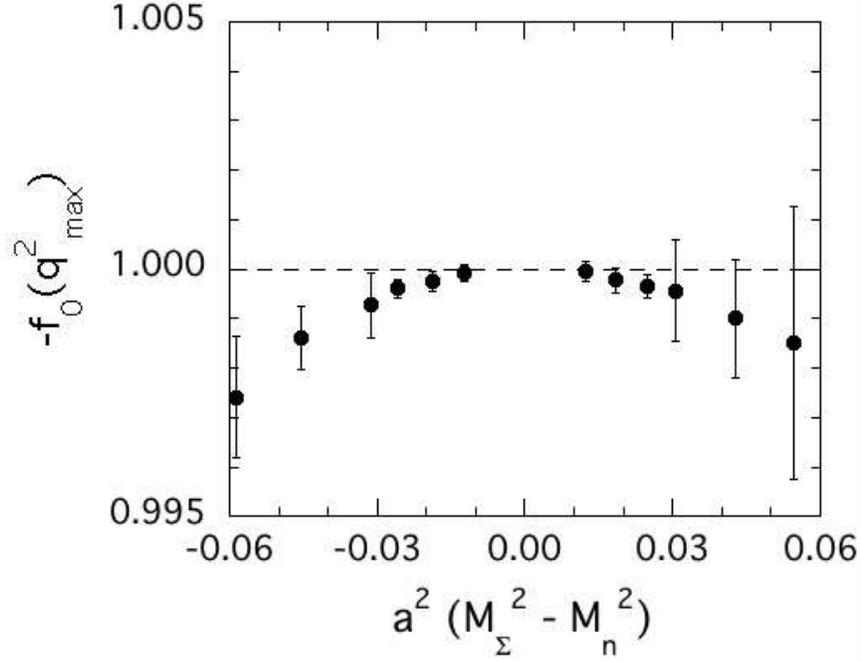}} \vspace{-0.50cm}
\end{center}
\caption{\it Results for $f_0(q^2_{max})$, for hyperon $\Sigma^- \to n \ell \nu$
decays, versus $a^2 (M_\Sigma^2 - M_n^2)$.}
\label{fig:f0(qmax)}
\end{figure}

The value of $f_0(q^2)$ at $q_{\rm max}^2 = (M_{\Sigma} - M_n)^2$ can be 
extracted using the double ratio (\ref{eq:fnal1}), which now reads as
\be 
    R_0(t_x, t_y)_{\overrightarrow{\stackrel{\mbox{\tiny $t_x \to \infty$}}
    {\mbox{\tiny $(t_y-t_x) \to \infty$}}}} \frac{\langle n | \bar{s} 
    \gamma_0 u | \Sigma \rangle \langle \Sigma | \bar{u} \gamma_0 s | 
    n \rangle}{\langle \Sigma | \bar{s} \gamma_0 s | \Sigma \rangle \langle n 
    | \bar{u} \gamma_0 u | n \rangle} = [f_0(q_{\rm max}^2)]^2 ~ .
    \label{eq:fnal3}
\ee
The results obtained for $f_0(q^2_{max})$ are shown in Fig.~\ref{fig:f0(qmax)}, 
where the high precision reached ($\lesssim 0.1$\%) can be appreciated.

\begin{figure}[t]
\begin{center}
\resizebox{0.850\textwidth}{!}{
\includegraphics{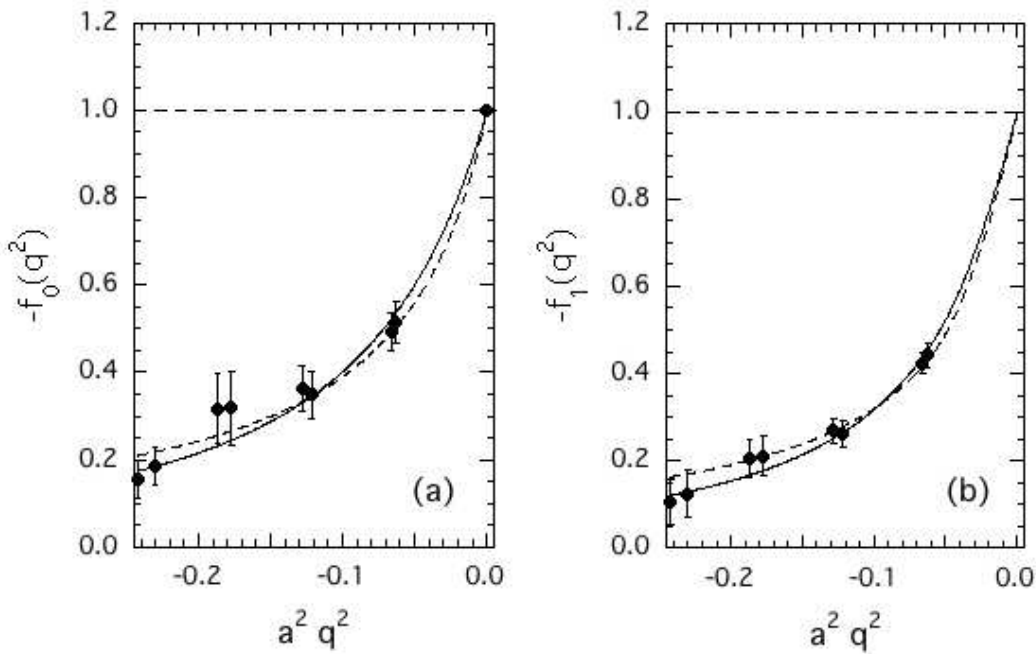}} \vspace{-0.50cm}
\end{center}
\caption{\it Results for $f_0(q^2)$ (a) and $f_1(q^2)$ (b) versus $a^2 q^2$. 
The dashed and solid lines represent respectively a monopole and a dipole fit 
to the lattice  data, [see Eq.~(\ref{eq:curves})].}
\label{fig:f0f1(q2)}
\end{figure}

Given the high accuracy reached at $q^2_{max}$ as well as the closeness of the 
values of $q_{\rm max}^2$ to $q^2 = 0$, it is enough to study the 
$q^2$-dependence of $f_{0, 1}(q^2)$ with an accuracy of about $10 \div 20$\% 
in order to reach the percent precision on $f_0(0)$. As in the case of 
mesons, the standard form factor analysis provides values of $f_1(q^2)$ quite 
well determined, whereas for $f_0(q^2)$ one has to resort to the double ratios 
(\ref{eq:fnal2}), which give access to the quantity $f_0(q^2)/f_1(q^2)$. In 
Fig.~\ref{fig:f0f1(q2)} we present the values of $f_0(q^2)$ and $f_1(q^2)$ 
obtained for a specific combination of quark masses. The points are paired 
because both $\Sigma^- \to n$ and $n \to \Sigma^-$ transitions are considered. 

In order to get the value $f_0(0) = f_1(0)$, the results for $f_0(q^2)$ and 
$f_1(q^2)$ are fitted by using either a monopole or a dipole functional form,
\be
   F^{(mon.)}(q^2) = \frac{A}{1 - q^2 / B} \qquad , \qquad  
   F^{(dip.)}(q^2) = \frac{C}{(1 - q^2 / D)^2} ~ , ~~
\label{eq:curves}
\ee
which nicely describe the lattice data, see Fig.~\ref{fig:f0f1(q2)}. In the case
of the vector form factor, the dipole parameter $\sqrt{D}$ agrees with the $K^*$
meson mass within $\simeq 15$\% accuracy. It should be mentioned that, from a 
theoretical point of view, the dipole parameterization of the form factors is 
only meant to represent, in an effective way, the contribution of several 
resonances. It is found that a dipole behaviour describes well the results of 
the electroproduction and neutrino experiments for $\Delta S =0$ transitions.
Thus, it might also provide a good description of the data in the $\Delta S = 1$
case~\cite{cabibbo,gaillard}.

\begin{figure}[t]
\begin{center}
\resizebox{0.70\textwidth}{!}{
\includegraphics{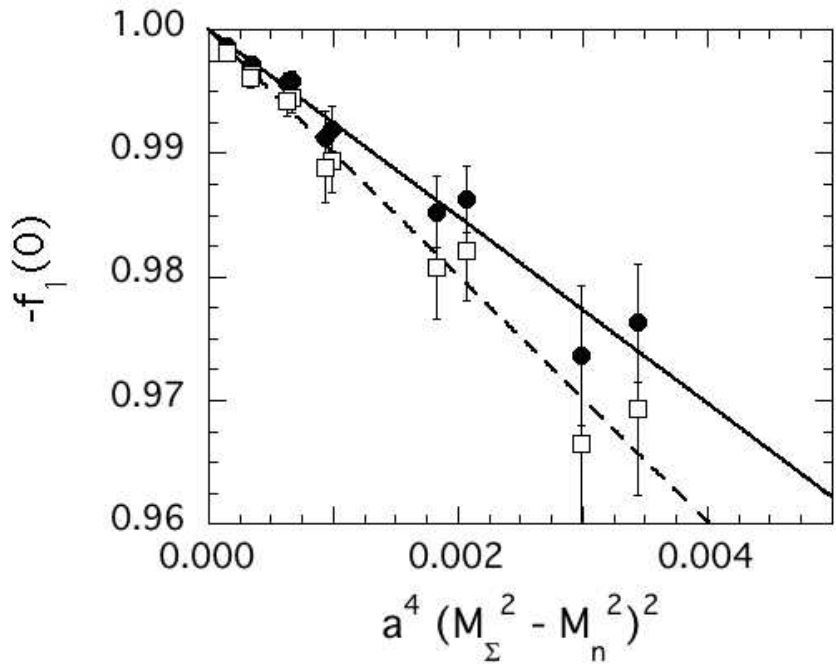}} \vspace{-0.50cm}
\end{center}
\caption{\it Results for $f_1(0)$ versus $a^4 (M_\Sigma^2 - M_n^2)^2$, obtained 
by performing a monopole (open squares) or a dipole (full circles) interpolation
in $q^2$. The dashed and solid lines are linear fits, according to the AG 
theorem \cite{ag}.}
\label{fig:f1(0)}
\end{figure}

In Fig.~\ref{fig:f1(0)} we collect the results obtained for $f_1(0)$. They 
clearly exhibit the nice linear dependence expected from the AG theorem. Thus 
$SU(3)$-breaking effects are resolved with a good precision even within the 
limited statistics used in Ref.~\cite{hyperons}.

As already pointed out in the Introduction, the evaluation of the leading chiral
correction $f_2$ for the baryon form factor $f_1(0)$ is more involved with 
respect to the case of $K_{\ell3}$ decays. We believe that the high-precision 
results obtained for hyperon decays encourage a more refined calculation of 
$f_2$, both in the quenched and unquenched theory, which would permit to control
the chiral extrapolation of $f_1(0)$ at a level of accuracy comparable to the 
one achieved for $f_+(0)$ in Ref.~\cite{our}. Work in this direction is in 
progress.

\section{Conclusions \label{sec:conclusions}}
We have presented quenched lattice studies of the $K \to \pi$ and $\Sigma \to n$
vector form factors at zero-momentum transfer, $f_+(0)$ and $f_1(0)$, 
respectively. Our calculations are the first ones obtained by using a 
non-perturbative method based only on QCD, except for the quenched 
approximation. The main goal is the determination of the SU(3)-breaking effects 
on $f_+(0)$ and $f_1(0)$, which is necessary to extract $|V_{us}|$ from 
$K_{\ell 3}$ and hyperon decays. In order to reach the required level of 
precision we have employed the double ratio method originally proposed in 
Ref.~\cite{FNAL} for the study of heavy-light form factors. We have found that 
this approach yields a determination of the scalar form factor $f_0(q^2)$ at 
$q^2 = q_{\rm max}^2$ with a statistical uncertainty well below $1$\% for both 
mesons and baryons.

A second step is the extrapolation of the scalar form factor to $q^2=0$. This 
has been performed by fitting the accurate results obtained using suitable 
double ratios of three-point correlation functions. The values of $f_+(0)$ and 
$f_1(0)$ obtained in this way are determined within the percent level of 
precision, which is the one required for a significant determination of 
$|V_{us}|$.

In the case of $K_{\ell3}$ decays, the leading chiral artifacts of the quenched 
approximation, represented by $f_2^q$, have been corrected for by means of an 
analytic calculation in quenched ChPT. After this subtraction, the lattice 
results are smoothly extrapolated to the physical meson masses, and we obtain
\be
   f_+^{K^0\pi^-}(0) = 0.960 \pm 0.005_{\rm stat} \pm 0.007_{\rm syst} ~ ,
   \label{eq:final}
\ee
where the systematic error does not include an estimate of quenched effects 
beyond ${\cal{O}}(p^4)$. Removing this error represents one of the major goal of
future lattice studies of $K_{\ell 3}$ decays. In this paper we have addressed 
in some details the impact of the lattice result on the determination of 
$|V_{us}|$. By considering only the new experimental data on $K_{\ell 3}$ decay 
widths from BNL-E865 \cite{E865}, KTeV \cite{KTeV}, NA48 \cite{NA48} and KLOE 
\cite{KLOE}, we find $|V_{us}| = 0.2250 \pm 0.0021$, in excellent agreement with
CKM unitarity. The results of a preliminary lattice study of hyperon decays, 
based on a similar strategy, have been also presented. 

\section*{Acknowledgements} The work of G.I.~and F.M.~is partially supported by 
IHP-RTN, EC contract No.~HPRN-CT-2002-00311 (EURIDICE).


\begin{thebibliography}{99}

\bibitem{CKM} N. Cabibbo, Phys.\ Rev.\ Lett.\  \textbf{10} (1963) 531. M. 
Kobayashi and T. Maskawa, Prog.\ Theor.\ Phys.\ \textbf{49} (1973) 652.

\bibitem{cabibbo} N. Cabibbo, E.C. Swallow and R. Winston, Phys.\ Rev.\ Lett.\ 
\textbf{92} (2004) 251803 [hep-ph/0307214]; Ann.\ Rev.\ Nucl.\ Part.\ Sci.\ 
\textbf{53} (2003) 39 [hep-ph/0307298].

\bibitem{milc} C.~Aubin {\it et al.} [MILC Collaboration], hep-lat/0407028.

\bibitem{marciano-vus} W.~J.~Marciano, hep-ph/0402299.

\bibitem{tau} E.~Gamiz, M.~Jamin, A.~Pich, J.~Prades and F.~Schwab, 
hep-ph/0408044.

\bibitem{PDG} PDG: S. Eidelmann {\em et al.}, Phys.\ Lett.\ \textbf{B592} 
(2004) 1.

\bibitem{ag} M. Ademollo and R. Gatto, Phys.\ Rev.\ Lett.\ \textbf{13} (1964) 
264.

\bibitem{LR} H. Leutwyler and M. Roos, Z.\ Phys.\ \textbf{C25} (1984) 91.

\bibitem{BT} J. Bijnens and P. Talavera, Nucl.\ Phys.\ \textbf{B669} (2003) 341
[hep-ph/0303103]. 

\bibitem{post} P. Post and K. Schilcher, Eur.\ Phys.\ J.\ \textbf{C25} (2002) 
427 [hep-ph/0112352].

\bibitem{CKNRT2} V. Cirigliano, H. Neufeld and H. Pichl, Eur.\ Phys.\ J.\ 
\textbf{C35} (2004) 53 [hep-ph/0401173].

\bibitem{jamin} M.~Jamin, J.~A.~Oller and A.~Pich, JHEP {\bf 0402} (2004) 047
[hep-ph/0401080].

\bibitem{anderson} J. Anderson and M.A. Luty, Phys.\ Rev.\ \textbf{D47} (1993) 
4975 [hep-ph/9301219].

\bibitem{manohar} E. Jenkins and A.V. Manohar, Phys.\ Lett.\ \textbf{B259} 
(1991) 353.

\bibitem{models} J.F. Donoghue, B.R. Holstein and S.W. Klimt, Phys.\ Rev.\ 
\textbf{D35} (1987) 934. F. Schlumpf, Phys.\ Rev.\ \textbf{D51} (1992) 2262. R. Flores-Mendieta,  
E. Jenkins and A.V. Manohar, Phys.\ Rev.\ \textbf{D58} (1998) 094028.  

\bibitem{our} D. Becirevic {\em et al.}, hep-ph/0403217. Accepted for
publication on Nucl. Phys. B.

\bibitem{dafneproc} V. Lubicz {\em et al.}, Proceedings of {\em DA$\Phi$NE 2004:
Physics at meson factories}, Laboratori Nazionali di Frascati (Italy), June 
7-11, 2004.

\bibitem{elbaproc} D. Becirevic {\em et al.}, Proceedings of {\em VIII 
International Conference on ``Electron-Nucleus Scattering"}, Marciana Marina 
(Italy), June 21-25, 2004.

\bibitem{latt04proc} D. Becirevic {\em et al.} [SPQ$_{\mathrm CD}$R 
Collaboration], Proceedings of {\em Lattice 2004}, Fermi National Accelerator
Laboratory, Batavia, Illinois (USA), June 21-26, 2004.

\bibitem{mesciaproc} F. Mescia, Proceedings of {\em ICHEP 2004}, Beijing
(China), August 16-22, 2004, hep-ph/0411097.

\bibitem{hyperons} D. Guadagnoli {\em et al.}, Proceedings of {\em Lattice 
2004}, Fermi National Accelerator Laboratory, Batavia, Illinois (USA), June 
21-26, 2004, hep-lat/0409048.

\bibitem{Cirigliano_old}
V.~Cirigliano  {\it et al.},
Eur.\ Phys.\ J.\ C {\bf 23} (2002) 121 [hep-ph/0110153].

\bibitem{Andre}
T.~C.~Andre, hep-ph/0406006.

\bibitem{E865} A. Sher {\it et al.} [E865 Collaboration], Phys.\ Rev.\ Lett.\ 
\textbf{91} (2003) 261802 and hep-ex/0307053.

\bibitem{KTeV}
T.~Alexopoulos {\it et al.}  [KTeV Collaboration], hep-ex/0406001.

\bibitem{NA48} A.~Lai {\it et al.}  [NA48 Collaboration], hep-ex/0410059;
L.~Litov, talk presented at {\em ICHEP \\ 2004} (Beijing, China, August 16-22,
2004), \\ {\tt http://www.ihep.ac.cn/data/ichep04/ppt/8\_cp/8-0432-litov-l.ppt}

\bibitem{KLOE} 
P. Franzini, invited talk at {\em PIC 2004} (Boston, USA, 27-29 Jun 2004), 
hep-ex/0408150; 
M. Antonelli, talk presented at {\em ICHEP '04} (Beijing, China, August 16-22 
2004), \\
{\tt http://www.ihep.ac.cn/data/ichep04/ppt/8\_cp/8-0811-antonelli-m.pdf.}

\bibitem{KTeV_slope}
T.~Alexopoulos {\it et al.}  [KTeV Collaboration], hep-ex/0406003.

\bibitem{marciano}
A.~Czarnecki, W.~J.~Marciano and A.~Sirlin, hep-ph/0406324.

\bibitem{GI}
G.~Isidori, eConf {\bf C0304052} (2003) WG601 [hep-ph/0311044].

\bibitem{FNAL} S. Hashimoto {\em et al.}, Phys.\ Rev.\ \textbf{D61} (2000) 
014502 [hep-ph/9906376].

\bibitem{ISTRA+}
O.~P.~Yushchenko {\it et al.}, Phys.\ Lett.\ B {\bf 581}, 31 (2004) 
[hep-ex/0312004]; Phys.\ Lett.\ B {\bf 589}, 111 (2004) [hep-ex/0404030].

\bibitem{NA48_slope}
A.~Lai {\it et al.}  [NA48 Collaboration], hep-ex/0410065.

\bibitem{gaillard} J.-M. Gaillard and G. Sauvage, Ann.\ Rev.\ Nucl.\ Part.\ 
Sci.\ \textbf{53} (1984) 34.

\end{thebibliography}
\end{document}